# PEP Analysis of Selective Decode and Forward Protocol over Keyhole Fading

Ravi Shankar, Yamini Chandrakar, Radhika Sinha, Ritesh Kumar Mishra

Department of Electronics and Communication Engineering
National Institute of Technology, Patna– 800005, INDIA
ravi.mrce@gmail.com

**ABSTRACT**

We provide a closed form upper bound formulation for the average pairwise-error probability (PEP) of selective decode and forward (SDF) cooperation protocol for a keyhole (pinhole) channel condition. We have employed orthogonal space-time block-code scheme (OSTBC) in conjunction with multi-antenna (MIMO) technology. We have used moment generating function (MGF) based approach for deriving the upper bound of PEP. PEP expression provides information regarding the performance of the wireless system with respect to the channel conditions. We have included simulation results which confirm the analytical results of our proposed upper bound. Simulation results show that due to keyhole effect performance of wireless system degrades.

## INTRODUCTION

Cooperative communications [1] - [2] enhances the performance of wireless systems, in which, user's co-ordinate among themselves to facilitate the transfer of information to each other. Two mostly used relay protocols are Amplify-and-Forward [3] - [4] and decode-and-Forward [5] - [6]. In case of AF relaying protocol source first sends the data to both relay as well as destination node. Relay-node amplifies the received signal followed by retransmission of a signal to the destination-node. However, in case of decode and forward protocol, relay-node decodes the signal it gets from the source and after decoding the signal it forwards the signal to the destination node. The problem associated with simple decodes and forward protocol is the transmission of erroneously decoded signal to the destination node. In order to overcome this problem, we are using SDF [7] relaying. In SDF relaying, relay only forwards the signal to the destination when it decodes correctly otherwise it will remain idle. In paper [8] -[9], the authors have analyzed the symbol error rate (SER) performance of the DF protocol over Rayleigh fading channel as well as Nakagami-m fading channel conditions. In [10], the author performed a PEP performance analysis of SDF over Rayleigh fading channel. In [11], the authors provide an enlightenment on the PEP performance of C(0) decode-and-forward cooperation protocol over Nakagami-m fading channel condition. To the best of our knowledge, PEP analysis has not been done for SDF over keyhole channel [12] as well as over keyhole Nakagami-m fading channels [13]. To this end, we have done PEP performance analysis of SDF over general keyhole channel condition and also keyhole Nakagami-m fading channel condition.

This paper is organized as follows: Section I contains the introduction of the system model and keyhole channel. The PEP close form expression is derived in section II for Gaussian keyhole channel. In section III, PEP expression is presented for keyhole Nakagami-m fading channel. In section IV simulation results are given to confirm our analytical results. In section V, the conclusion is given.

*Notations:* The mathematical notation $B^T$ is the transpose of the matrix. $B^H$ is the transpose of hermitian of the matrix. $|b|$ indicate magnitude of $a$ which is a complex number. $\|B\|_F$ indicate the matrix Frobenius norm of $B$. $E(.)$ indicate the expectation operation. $_pF_q(a_1, a_2, ..., a_p; b_1, b_2, ..., b_q; x)$ is generalized hypergeometric function. Also $Q(z)$ is the Gaussian $Q$ function. The standard Gaussian r.v. $Z$ whose mean is zero and variance is one, written as $N(0,1)$ has right tail probability as $P(Z \geq z)$. Also Notation $\mathbb{C}^{N \times T}$ indicate $N \times T$ matrices over the complex field $\mathbb{C}$.

## I. SYSTEM MODEL

### A. Moment generating function of Gaussian keyhole channel

Keyhole (pinhole) channel condition arises in high scattering around the receiver and transmitter. Consider a screen with a small barrier (pinhole) (e.g. tunnel) between the $N_T$ transmitting node and $N_R$ receiving node. The signal can go to the receiver only by propagating through this pinhole. This barrier causes a reduction in the rank of the channel coefficient matrix and degradation in performance of the wireless communication [14] system. The channel matrix can be modelled by considering that channel fading coefficients of H matrix are a product of two un-correlated Gaussian random variables. In Fig. 1 we have given the diagram of the keyhole fading channel.

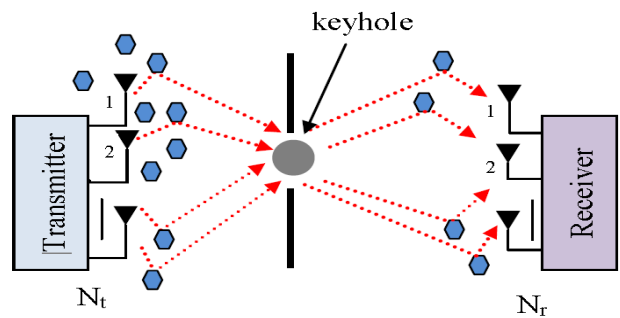

**Fig. 1 Keyhole Propagation Model**

Let $\beta \in \mathbb{C}^{N_R \times 1} = [\beta_1, \beta_2, \beta_3, \ldots, \beta_{N_R}]^T$ and $\alpha \in \mathbb{C}^{N_T \times 1} = [\alpha_1, \alpha_2, \alpha_3, \ldots, \alpha_{N_T}]^T$ Where $\alpha_j$ and $\beta_i$ are zero mean circular shift complex $\mathbb{CN}(0,1)$ Gaussian random variable [15], presenting scattering at the receiver and the transmitter respectively. The channel matrix is given as, $H = \beta \alpha^T$. Rank of $H$ is 1 i.e. due to keyhole effect the rank of the matrix decreases. Set, $Y = \|H\|^2 = \|\alpha\|^2 \|\beta\|^2 = UV$, where $U$ and $V$ are chi-square random variable with $2*N_R$ and $2*N_T$ degree of freedom. The moment generating function of Y is given below [16],

$$M_Y(s) = {}_2F_0(N_T, N_R; ; -s) \qquad (1)$$

### B. Nakagami-m keyhole fading channel

Let us consider a non-line of sight (N-LOS), local scattering channel model at both transmit (TX) and receive (RX) sides for outdoor transmission. Assuming scatters are close to either transmit (TX) or receive array, the channel matrix $H_{keyhole}$ for multiple-input multiple-output channels is given by,

$$H_{keyhole} = \beta \alpha^T \qquad (2)$$

Where, $\beta = \begin{bmatrix} \beta_1 e^{\mathbf{j}\theta_1} \\ \beta_2 e^{\mathbf{j}\theta_2} \\ \cdot \\ \cdot \\ \beta_{N_R} e^{\mathbf{j}\theta_{N_R}} \end{bmatrix}$

and $\alpha = \begin{pmatrix} \alpha_1 e^{\mathbf{j}\phi_1} & \alpha_2 e^{\mathbf{j}\phi_2} & \ldots & \alpha_{N_T} e^{\mathbf{j}\phi_{N_T}} \end{pmatrix}^T$

$$H_{keyhole} = \begin{pmatrix} \beta_1 e^{\mathbf{j}\theta_1} \\ \beta_2 e^{\mathbf{j}\theta_2} \\ \vdots \\ \beta_{N_R} e^{\mathbf{j}\theta_{N_R}} \end{pmatrix} \begin{pmatrix} \alpha_1 e^{\mathbf{j}\phi_1} & \alpha_2 e^{\mathbf{j}\phi_2} & \ldots & \alpha_{N_T} e^{\mathbf{j}\phi_{N_T}} \end{pmatrix}$$

$$= \begin{bmatrix} \alpha_1 \beta_1 e^{\mathbf{j}(\theta_1+\phi_1)} & \alpha_2 \beta_1 e^{\mathbf{j}(\theta_2+\phi_1)} & . & . & \alpha_{N_T}\beta_1 e^{\mathbf{j}(\theta_{N_T}+\phi_1)} \\ \alpha_1 \beta_2 e^{\mathbf{j}(\theta_1+\phi_2)} & \alpha_2 \beta_2 e^{\mathbf{j}(\theta_2+\phi_2)} & . & . & \alpha_{N_T}\beta_2 e^{\mathbf{j}(\theta_{N_T}+\phi_2)} \\ . & . & . & . & . \\ . & . & . & . & . \\ \alpha_1 \beta_{N_R} e^{\mathbf{j}(\theta_1+\phi_{N_T})} & \alpha_2 \beta_{N_R} e^{\mathbf{j}(\theta_2+\phi_{N_T})} & . & . & \alpha_{N_T}\beta_{N_R} e^{\mathbf{j}(\theta_{N_R}+\phi_{N_T})} \end{bmatrix}$$

In equation (3), $\{\alpha_j e^{\mathbf{j}\phi_j}\}_{j=1}^{N_T}$ and $\{\beta_i e^{\mathbf{j}\theta_i}\}_{i=1}^{N_R}$ denote the rich transmit (Tx) and receive (Rx) array respectively and bold letter $\mathbf{j} = \sqrt{-1}$. Each entry of $H$ is denoted as $h_{ij}$, from the $j^{th}$ antenna to the $i^{th}$ transmit antenna. $(\alpha_j)_{j=1}^{N_t}$ and $(\beta_i)_{i=1}^{N_r}$ are independent and identically distributed $(i.i.d.)$ Nakagami-m random variables with fading parameter $m_T$ and $m_R$ respectively, distributed as:

$$p_{\alpha_j}(\alpha) = \frac{2}{\overline{|(m_T)}} \left(\frac{m_T}{\Omega_T}\right)^{m_T} \alpha^{2m_T - 1} e^{-\frac{m_T \alpha^2}{\Omega_T}}$$

$\alpha \geq 0, m_T \geq 0.50, j = 1,2,3\ldots\ldots N_T$

$$p_{\beta_i}(\beta) = \frac{2}{\overline{|(m_R)}} \left(\frac{m_R}{\Omega_R}\right)^{m_R} \beta^{2m_R - 1} e^{-\frac{m_R \beta^2}{\Omega_R}},$$

$\beta \geq 0, m_R \geq 0.50, i = 1,2,3\ldots\ldots N_R$,

Where $\Omega_T = E[\alpha_j^2], \Omega_R = E[\beta_i^2]$ and $\overline{|(.)}$ denote the gamma function. $\{\phi_j\}_{j=1}^{N_T}$ and $\{\theta_i\}_{i=1}^{N_R}$ denote the independent and uniformly distribution over $[0, 2\pi]$. Regeneration of captured energy takes place in keyhole Nakagami-m fading channel.

### C. Moment generating function (MGF) of keyhole Nakagami-m Fading channel

Let $U = \sum_{j=1}^{N_T} \alpha_j^2$ and $V = \sum_{i=1}^{N_R} \beta_i^2$, the $L_2$ (Frobenius) norm of $H$ is given by,

$$\| H_{keyhole} \|_F^2 = \sum_{i=1}^{N_R}\sum_{j=1}^{N_T} \alpha_j^2 \beta_i^2 = \sum_{j=1}^{N_T} \alpha_j^2 \sum_{i=1}^{N_R} \beta_i^2 = U.V$$

As, $\alpha_j$ and $\beta_i$ are independent and identically distributed $(i.i.d.)$ Nakagami-m random variables $\alpha_j^2$ and $\beta_i^2$ should be Gamma distribution, i.e. $\alpha_j^2 \sim \gamma\left(\frac{\Omega_T}{m_T}, m_T\right)$ and $\beta_i^2 \sim \gamma\left(\frac{\Omega_R}{m_R}, m_R\right)$ as given in [17], for unit channel variance, the moment generating function of squared Frobenius norm $\| H_{keyhole} \|_F^2$ is given by

$$\| H_{keyhole} \|_F^2 = {}_2F_0\left(m_T N_T, m_R N_R; ; -\frac{s}{m_T m_R}\right) \qquad (4)$$

## II. PEP EXPRESSION FOR GAUSSIAN KEYHOLE CHANNEL

Let $C$ denotes the set of code word matrices, $C = \{X_j\}$ and $X_j \in \mathbb{C}^{N_S * T}$ for $T$ timeslot $1 \leq j \leq |C|$ denotes each code word of set $C$ where $|C|$ represent the cardinality of the codeword set. $N_S$ is number of antennas at source. The System model for SDF is given in Fig.2

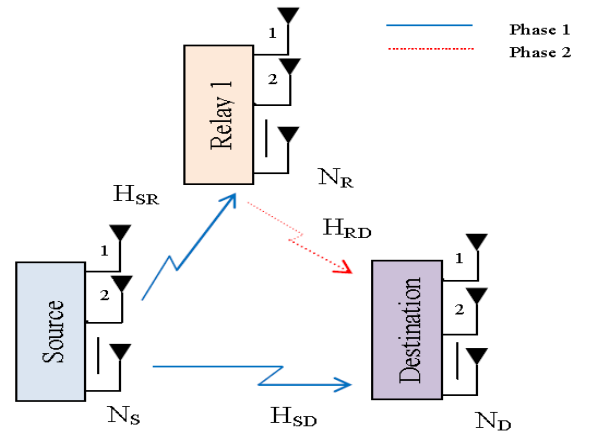

**Fig. 2 System Model of SDF**

In phase 1, Direct source to destination transmission as well as source to relay transmission takes place. Phase 2 only occur when relay correctly decode the received codeword and relay to destination transmission takes place.

The instantaneous expression of PEP that the relay erroneously decodes $X_j$ from a codeword matrix $X_0$ is given as [18],

$$P(X_0 \to X_j | \beta_l \alpha_J) = Q\left(\sqrt{\frac{P_0}{2N_S\eta_0}trace\left(H_{SR}(X_0-X_j)(X_0-X_j)^H H_{SR}^H\right)}\right)$$

$$= \frac{1}{\pi}\int_0^{\pi/2} \exp\left(-\frac{P_0}{2\times 2N_S\eta_0 \sin^2\theta}trace\left(H_{SR}(X_0-X_j)(X_0-X_j)^H H_{SR}^H\right)\right) \quad (5)$$

Here $P_0$ is total source power, $\eta_0$ as noise at source. $H_{SR}$ is gaussian keyhole channel. As $(X_0-X_j)(X_0-X_j)^H$ is positive definite, hence it admits a singular value decomposition (SVD) given as,

$$(X_0-X_j)(X_0-X_j)^H = U_j \Lambda_j U_j^H \quad (6)$$

Using equation (6) in equation (5), we get,

$$P(X_0 \to X_j | \beta_l \alpha_J)$$
$$= \frac{1}{\pi}\int_0^{\pi/2} \exp\left(-\frac{P_0}{4N_S\eta_0 \sin^2\theta} trace\left(\beta\left(\alpha^T U_j \Lambda_j U_J^H (\alpha^T)^H\right)\beta^H\right)\right) \quad (7)$$

Setting $\gamma = \alpha^T U = [\gamma_1 \gamma_2 \ldots \gamma_{N_S}]$. $U_j$ being unitary matrix Hence $|\gamma|^2 = |\alpha|^2$. As we get,

$$\alpha^T U_j \Lambda_j U_j^H (\alpha^T)^H = \gamma^T \Lambda_j (\gamma^T)^H$$

$$\gamma^T \Lambda_j (\gamma^T)^H = \lambda_1 |\gamma_1|^2 + \lambda_2 |\gamma_2|^2 + \ldots + \lambda_{N_S} |\gamma_{N_S}|^2$$

Where $\lambda_i$ are positive singular values and $i \geq 0$. As, singular values are in decreasing order hence $\lambda_1 \geq \lambda_2 \geq \ldots \geq \lambda_{N_S}$.

Also, we can further simplify $\alpha^T U_j \Lambda_j U_j^H (\alpha^T)^H$ as,

$$\alpha^T U_j \Lambda_j U_j^H (\alpha^T)^H$$
$$= \lambda_{N_S}\left(\frac{\lambda_1}{\lambda_{N_S}}|\gamma_1|^2 + \frac{\lambda_2}{\lambda_{N_S}}|\gamma_2|^2 + \ldots + |\gamma_{N_S}|^2\right)$$
$$\geq \lambda_{N_S}\left(|\gamma_1|^2 + |\gamma_2|^2 + \ldots + |\gamma_{N_S}|^2\right)$$
$$= \lambda_{N_S}(|\gamma|^2)$$
$$= \lambda_{\min}^{X_0,X_j} |\alpha|^2 \quad (8)$$

Putting the value of equation (8) in equation (7), we get,

$$P_{S \to R}(X_0 \to X_j | \beta_l \alpha_J) \leq \frac{1}{\pi}\int_0^{\pi/2} \exp\left(-\frac{P_0 \lambda_{\min}^{X_0,X_j} \|\alpha\|^2 \|\beta\|^2}{4N_S\eta_0 \sin^2\theta}\right)d\theta$$

$$= \frac{1}{\pi}\int_0^{\pi/2} \exp\left(-\frac{P_0 \lambda_{\min}^{X_0,X_j} \|H_{SR}\|_F^2}{4N_S\eta_0 \sin^2\theta}\right)d\theta \quad (9)$$

Taking the expected value of expected value of equation (9) we get,

$$P_{S \to R}(X_0 \to X_j) = E\left[\frac{1}{\pi}\int_0^{\pi/2} \exp\left(-\frac{P_0 \lambda_{\min}^{X_0,X_j} \|H_{SR}\|_F^2}{4N_S\eta_0 \sin^2\theta}\right)d\theta\right]$$

From moment generating function given in equation (1) we get,

$$P_{S \to R}(X_0 \to X_j) = \frac{1}{\pi}\int_0^{\pi/2} {}_2F_0\left(N_S, N_R;; -\frac{P_0 \lambda_{\min}^{X_0,X_j}}{4N_S\eta_0 \sin^2\theta}\right)d\theta$$

$\lambda_{\min}^{X_0,X_j} \Rightarrow$ Minimum singular value.

$$P_{S \to R}(X_0 \to X_j) = \frac{1}{\pi}\int_0^{\pi/2} {}_2F_0\left(N_S, N_R;; -\frac{P_0 \lambda_{\min}^{X_0,X_j}}{4N_S\eta_0 \sin^2\theta}\right)d\theta$$

Upper bound expression for the total PEP is given by,

$$P_{S \to R} \leq \sum_{X_j \in C, X_j \neq X_0} P_{S \to R}(X_0 \to X_j)$$

$$= \sum_{i=1}^{|c|}\left[\frac{1}{\pi}\int_0^{\pi/2} {}_2F_0\left(N_S, N_R;; -\frac{P_0 \lambda_{\min}^{X_0,X_j}}{4N_S\eta_0 \sin^2\theta}\right)d\theta\right] \quad (10)$$

Average PEP of error event, $X_0 \to X_i$ corresponding to the direct source to destination transmission conditioned on channel matrix $\tilde{H}_{SD}$.

$$P_{S \to D}(X_0 \to X_i) = \frac{1}{\pi}\int_0^{\pi/2} {}_2F_0\left(N_S, N_D;; -\frac{P_0 \lambda_{\min}^{X_0,X_i}}{4N_S\eta_0 \sin^2\theta}\right)d\theta \quad (11)$$

Where $\tilde{H}_{SD}$ denote the channel fading matrix for source to destination $\tilde{H}_{SD} = H_{SD}U_i$. Let $h_{l_n}^{(SD)}$, $h_{l_n}^{(RD)}$ denote the Nakagami-m fading channel coefficients corresponding to source to destination and relay to destination channel matrices $\tilde{H}_{RD} = H_{RD}U_i$ and $\tilde{H}_{SD} = H_{SD}U_i$ respectively. In case when the relay node correctly decodes the signal obtained from the source node, we will write the expression of PEP conditioned on $\tilde{H}_{RD}$ and $\tilde{H}_{SD}$ as, in equation (12) where $P_1$ is relay power.

Net PEP of error event $X_0 \to X_i$ is given below;

$$P_e(X_0 \to X_i) = P_{S \to D}(X_0 \to X_i) \times P_{S \to R} + P_{S \to D, R \to D}(X_0 \to X_i) \times (1-P_{S \to R}) \quad (13)$$

Considering $(1-P_{S \to R}) \approx 1$ at high SNR and using equation (10) and (12) in equation (13), we can write union bound expressions for the PEP as;

$$P_e \leq \sum_{i=1}^{|c|} P_e(X_0 \to X_i)$$

Hence end to end PEP at destination is expressed in equation (14).

### III. PEP EXPRESSION FOR KEYHOLE NAKAGAMI-M CHANNEL

Using the same procedure and by using the moment generating function approach as given in equation (34) in paper [13], we can write the PEP upper bound expression as expressed in equation (15) where $m_{SD}$ is fading parameter at source-destination and $m_{SR}$ is fading parameter at between source-relay node and $m_{RD}$ is the fading parameter of the relay-destination node. $\Omega_{SD}$, $\Omega_{SR}$ and $\Omega_{RD}$ are the channel variances of the source-destination, source-relay and relay-destination respectively.

## IV. SIMULATION RESULTS AND DISCUSSION

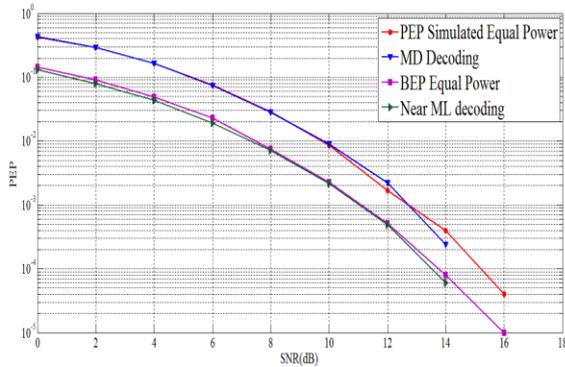

**Fig. 3 PEP vs SNR(dB) for Keyhole Gaussian Channel with var_sd=var_sr=var_rd=1**

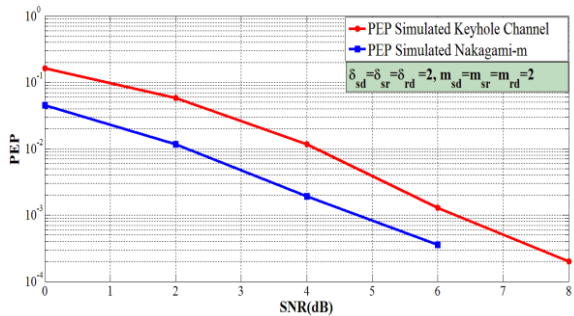

**Fig. 4 PEP versus Signal to Noise Ratio (dB) for Nakagami-m and keyhole Nakagami-m fading channel**

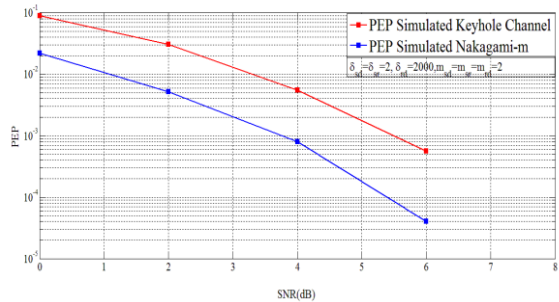

**Fig. 5 PEP versus Signal to Noise Ratio (dB) for Nakagami-m and keyhole Nakagami-m fading channel**

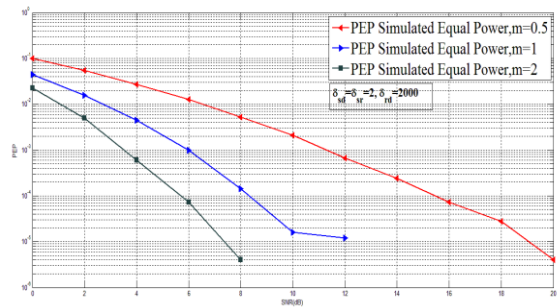

**Fig. 6 PEP versus Signal to Noise Ratio (dB) for keyhole Nakagami-m fading channel with variation in m fading parameter**

In Fig. 3 we have given the PEP performance for various types of decoding schemes. From Figure 3 we have shown that bit error probability performance is very near to the performance of maximum likelihood decoding schemes. The performance of the minimum distance decoding scheme is worst as compared to the near maximum likelihood and PEP performance. This verifies out theoretical results because in case of minimum distance decoding relay might forward the erroneous signal to the destination, so its performance is bad as compared to PEP.

In Fig. 4 we have taken $\Omega_{SD} = \Omega_{SR} = \Omega_{RD} = 2$ and $m_{SD} = m_{SR} = m_{RD} = 2$. In this figure we have compared the PEP performance between the Nakagami-m channel and keyhole Nakagami-m fading channel. We can show that PEP increases in case of keyhole Nakagami-m channel, it means performance decreases in Keyhole channel. This verifies our analytical result that due to rank deficiency the performance of the keyhole Nakagami-m fading channel is worst as compared to Nakagami-m fading channel. In Fig. 5 we have repeated the same comparison, but for $\Omega_{SD} = \Omega_{SR} = \Omega_{rD} = 2000$. In Fig. 5 also we have shown that the performance of keyhole channel is worst as compared to Nakagami-m fading channel. In Fig. 6 we have shown the performance of PEP in variation with the $m$ fading parameter. For higher values of $m$ we obtain lower PEP.

## V. CONCLUSION

We have formulated the PEP expression for the keyhole considering Gaussian channel as well as Keyhole Nakagami-m fading channel. We have employed orthogonal space-time block coding (OSTBC) in conjunction with multi-antenna (MIMO) technology. We have conducted simulation which affirms the analytical results of our proposed PEP expression for Keyhole channel. We have shown that due to keyhole the performance of the system decreases. Simulation results show that due to rank deficiency keyhole produces worst performance as compared to Nakagami-m fading channel.

$$P_{S \to D, R \to D}(X_0 \to X_i) = \frac{1}{\pi} \int_0^{\pi/2} \left[ {}_2F_0\left(N_S, N_D;;-\frac{P_0 \lambda_{\min}^{X_0,X_i}}{4N_S \eta_0 \sin^2 \theta}\right) \times {}_2F_0\left(N_R, N_D;;-\frac{P_1 \lambda_{\min}^{X_0,X_i}}{4N_R \eta_0 \sin^2 \theta}\right) \right] d\theta \quad (12)$$

$$P_e \leq \sum_{i=1}^{|c|} \left[ \frac{1}{\pi} \int_0^{\pi/2} \left[ {}_2F_0\left(N_S, N_D;;-\frac{P_0 \lambda_{\min}^{X_0,X_i}}{4N_S \eta_0 \sin^2 \theta}\right) \right] d\theta \times \sum_{j=1}^{|c|} \frac{1}{\pi} \int_0^{\pi/2} \left[ {}_2F_0\left(N_S, N_R;;-\frac{P_0 \lambda_{\min}^{X_0,X_j}}{4N_S \eta_0 \sin^2 \theta}\right) \right] d\theta \\ + \frac{1}{\pi} \int_0^{\pi/2} \left[ {}_2F_0\left(N_S, N_D;;-\frac{P_0 \lambda_{\min}^{X_0,X_i}}{4N_S \eta_0 \sin^2 \theta}\right) \times {}_2F_0\left(N_R, N_D;;-\frac{P_1 \lambda_{\min}^{X_0,X_i}}{4N_R \eta_0 \sin^2 \theta}\right) \right] d\theta \right] \quad (14)$$

$$P_e \leq \sum_{i=1}^{|c|} \left[ \frac{1}{\pi} \int_0^{\pi/2} \left[ {}_2F_0\left(m_{SD}N_S, m_{SD}N_D;;-\frac{P_0 \lambda_{\min}^{X_0,X_i}\Omega_{SD}}{4N_S \eta_0 \sin^2 \theta}\right) \right] d\theta \times \sum_{j=1}^{|c|} \frac{1}{\pi} \int_0^{\pi/2} \left[ {}_2F_0\left(m_{SR}N_S, m_{SR}N_R;;-\frac{P_0 \lambda_{\min}^{X_0,X_j}\Omega_{SR}}{4N_S \eta_0 \sin^2 \theta}\right) \right] d\theta \\ + \frac{1}{\pi} \int_0^{\pi/2} \left[ {}_2F_0\left(m_{SD}N_S, m_{SD}N_D;;-\frac{P_0 \lambda_{\min}^{X_0,X_i}\Omega_{SD}}{4N_S \eta_0 \sin^2 \theta}\right) \times {}_2F_0\left(m_{RD}N_R, m_{RD}N_D;;-\frac{P_1 \lambda_{\min}^{X_0,X_i}\Omega_{RD}}{4N_R \eta_0 \sin^2 \theta}\right) \right] d\theta \right] \quad (15)$$